# Data Likelihood of Active Fires Satellite Detection and Applications to Ignition Estimation and Data Assimilation

James Haley[1]; Angel Farguell Caus[2];

Adam K. Kochanski[3]; Sher Schranz[4]; Jan Mandel[5]*

[1] *University of Colorado Denver, Denver, Colorado, 0000-0001-5121-5456,*
*{james.haley@ucdenver.edu}*
[2] *University of Colorado Denver, Denver, Colorado, and Autonomous University of*
*Barcelona, Spain, 0000-0003-2986-3857, {angel.farguell@gmail.com}*
[3]*University of Utah, Salt Lake City, Utah, {adam.kochanski@utah.edu}*
[4]*CIRA, Colorado State University, Ft. Collins, Colorado {sher@rams.colostate.edu}*
[5]* *University of Colorado Denver, Denver, Colorado, 0000-0002-8489-5766,*
*{jan.mandel@ucdenver.edu}*

**Abstract**

Data likelihood of fire detection is the probability of the observed detection outcome given the state of the fire spread model. We derive fire detection likelihood of satellite data as a function of the fire arrival time on the model grid. The data likelihood is constructed by a combination of the burn model, the logistic regression of the active fires detections, and the Gaussian distribution of the geolocation error. The use of the data likelihood is then demonstrated by an estimation of the ignition point of a wildland fire by the maximization of the likelihood of MODIS and VIIRS data over multiple possible ignition points.

## 1. Introduction

Satellite-based sensors are a commonly used data source because of their large spatial coverage, but the mismatch of scales, geolocation errors, the probabilistic character of the fire detection, and missing data present a challenge. A practical resolution for fire behavior models is given by fuel data availability, typically about 30m, while the resolution of satellite-based fire detections is from 375m to 2km and up. Therefore, initializing a model directly by igniting entire detection squares results in blocky, fragmented fire shape. Furthermore, in coupled fire-atmosphere models, such ignitions generally result in numerical instabilities induced by the sudden heat release from the ignited pixel, as well fire state inconsistent with the atmospheric state at the ignition time. In the presented method, we use satellite data to improve the fire modelling in a statistical sense. The basic tool we use is data likelihood, which is defined as the probability of the fire detection outcome given the state of the fire-spread model.

Data likelihood is one of the techniques to evaluate a model state relative to data, and a basic ingredient of many data-driven simulation methods. Data likelihood is used e.g., in data assimilation to update importance weights in particle filters, or as a term in an objective function of an optimization method in maximum likelihood or maximum aposteriori probability (MAP) estimates.

Other, less formal pragmatic fitness functions are often used for model evaluation, see a survey in Filippi et al. (2013). Approaches used in wildland fire spread include evaluation of a model solution by the difference of burned area e.g., Brun et al. (2017), who update the solution by genetic algorithms; least squares, or, equivalently, Gaussian data likelihood of perimeter position with update by sequential Monte Carlo (e.g., Srivas et al. 2017), or by the ensemble Kalman filter (EnKF) in Rochoux et al. (2014); Gaussian likelihood of sensor data (e.g., Gu 2018; Xue et al. 2012); and the size of a spatial deformation needed to match two fires, with the spatial deformation mapping extending the state in the EnKF (Beezley and Mandel, 2008; Mandel et al. 2009). See also a survey of data assimilation for wildland fire spread modeling in Gollner et al. (2015, Sect. 4).

Approaches to wildland fire modeling using satellite data include periodic reinitialization from new detections (Coen and Schroeder 2013; Sá et al. 2014) and identification of ignition as the first active fires detection within a reported perimeter (Benali et al. 2016). However, because of the missing data, the statistical uncertainty of detection, the geolocation uncertainty, and the mismatch of scales between the fire model and the satellite sensor, direct use of satellite data at fuel map scale is of limited value and the data is better suitable to improve models in a statistical sense. Fire behavior models run on a mesh given by fuel data availability, typically with about 30m resolution and aligned with geographic coordinates and with time step of the order of seconds. The satellite fire detections are commonly used due the large spatial coverage, but the gap of scales and data errors present a challenge. The resolution of satellite-based sensors is much coarser, from 375m once a day for VIIRS (Schroeder et al. 2014), 1.1km twice a day for MODIS (Giglio et al. 2016), aligned with flight coordinates, to 2km every 5 minutes for GOES-16 (Koltunov et al. 2016). Geolocation error can be significant, e.g., 1.5km at $3\sigma$ for VIIRS (Sei, 2011). Pixels may be missing for various reasons such as clouds. While consumer-grade data provide only fire detection pixels, science-grade data make an important distinction between missing data and absence of fire detections.

Spatial statistical interpolation of the first detection time by kriging to obtain a continuous fire arrival time field was proposed by Veraverbeke et al. (2014). Sá et al. (2017) proposed a measure of spatial discrepancy between fire spread simulation and satellite data. The probability of fire detection in a sensor pixel, given the state of the fire and the properties of the surface in the pixel, was estimated in a validation study (Schroeder et al., 2008) by logistic regression. Mandel et al. (2014b) suggested a form of data likelihood for satellite active fires detection in a pixel and used it for data assimilation by a MAP estimate. The form of the data likelihood from Mandel et al. (2014b) was further motivated in Mandel et al. (2016b) by substituting the fire heat flux into the logistic regression from Schroeder et al. (2008), which explained the behavior of the likelihood function after the fire arrival time. This likelihood function was used to find a maximum likelihood estimate of an ignition point in Mandel et al. (2016a).

In this paper, we build a data likelihood function by adding a geolocation error to the construction in Mandel et al. (2016b), which combines a heat release model with logistic regression for the active fires detections. For a single pixel, we recover and justify the form of data likelihood function proposed in Mandel et al. (2014b, 2016b). We then demonstrate use of the data likelihood on identifying the ignition point on a realistic example.

We have used WRF-SFIRE (Mandel et al., 2009, 2011, 2014a) in the examples in this paper. WRF-SFIRE evolved from CAWFE (Clark et al. 2004) and it has been a part of WRF release as WRF-Fire since 2011 (Mandel et al. 2011; Coen et al 2013).

## 2. Data Likelihood of Active Fires Satellite Detection

The state of the fire-spread model is encoded as fire arrival time on a grid of locations of Earth surface. The data likelihood is obtained from the fire arrival time by substituting the heat release into the logistic sensitivity function and convolution of the result with a Gaussian kernel to account for the geolocation error (Figure 1). See Appendix A for mathematical details.

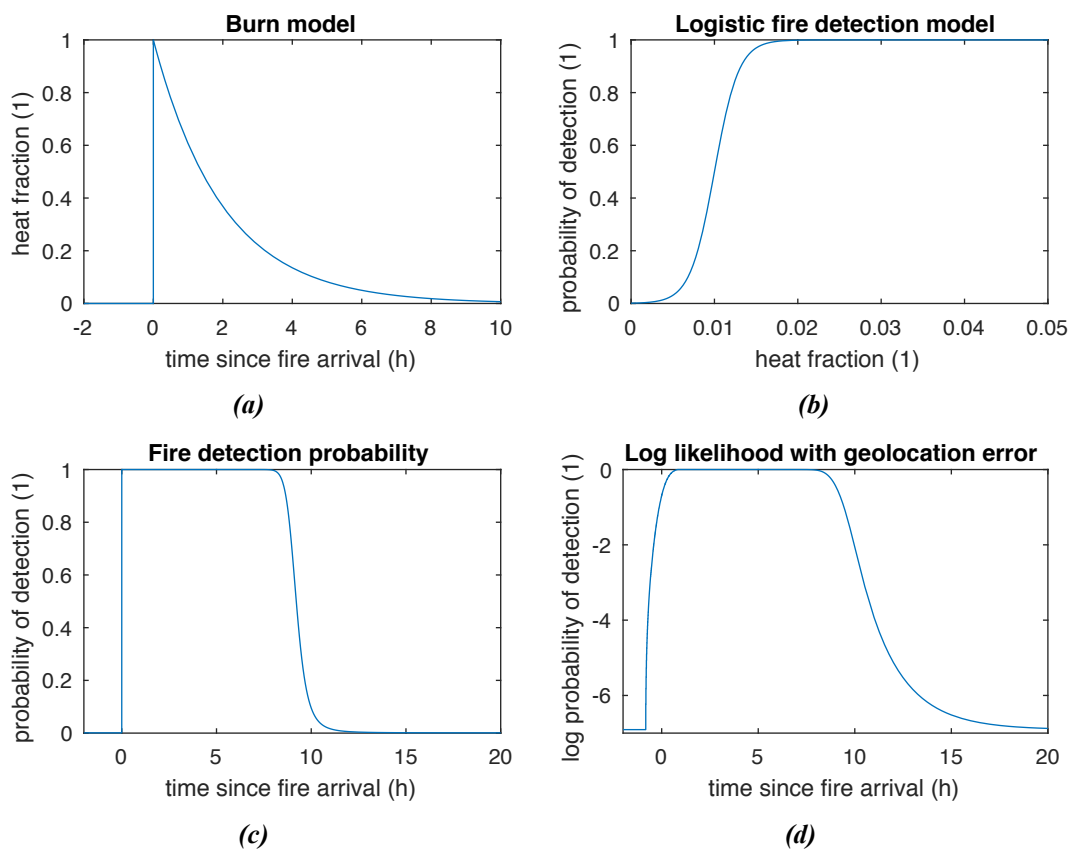

*Figure 1  (a) Heat release model with exponential decay. The maximum heat fraction of 1 drops to 1/e in the characteristic time of 2h. (b) Probability of fire detection as a logistic function of the fraction of maximum heat release, with 50% probability of detection at 1% heat fraction, and 0.1% false detection rate. (c) Probability of detection as a function of time elapsed since fire arrival, obtained by substituting the heat release fraction from (a) to the logistic curve (b). (d) Final log probability of detection from Eq. (7), obtained by convolution of (c) with a Gaussian kernel. The fire rate of spread was 1m/s and Gaussian geolocation error was σ=2000m , equal to a rounded sum of the diagonal of a 1.1km pixel plus the standard deviation of 0.5km of the geolocation error from VIIRS specifications (Sei, 2011).*

The log likelihood suggested in Mandel et al. (2014b, 2016b) was based on the pragmatic consideration that the probability of detection is close to one for some time after the fire arrival, with quadratic tails, and the leading edge is steeper than the trailing edge. Note that the log likelihood curve in Figure 1(d) has exactly this type of shape above the transition to the constant nonzero false detection rate.

## 3. Retrieving the Ignition Location and Time
### 3.1. A Simple Test

As a first test of whether the likelihood function can be used to retrieve the ignition location and time of a wildfire we made a simple experiment simulating an idealized fire over flat terrain with homogenous fuels and no winds. We picture the progression of the perimeter of such a fire to be originating from a point and growing outward like the wave caused by a stone dropped into a pool of still water. With $T$ representing the fire arrival time at some particular spatial point $(x,y)$, we model the progression of the fire as cone

$$T(x,y) = \sqrt{x^2 + y^2} \,.$$

Working with a spatial domain of 1000 units square, we simulate a fire with ignition point $(x,y) = (500,500)$ and ignition time $t = 30$ with the function

$$T(x,y) = \sqrt{(x-500)^2 + (y-500)^2} + 30.$$

We then simulated satellite fire detections at several points along the simulated fire perimeter corresponding to the level-curves of the function at time $t$=300. In a real-world setting, these detections may be the first information we have about a fire and in order to achieve the best simulation of it we need to have a good estimate of the ignition point. We find this estimate by running many simulations of the fire starting at locations and times near those of the satellite detections and then computing the log-likelihood of the simulated satellite detections according to equation (6). The largest log-likelihood obtained from the set of simulations gives us the best estimate of the location and time of the fire ignition.

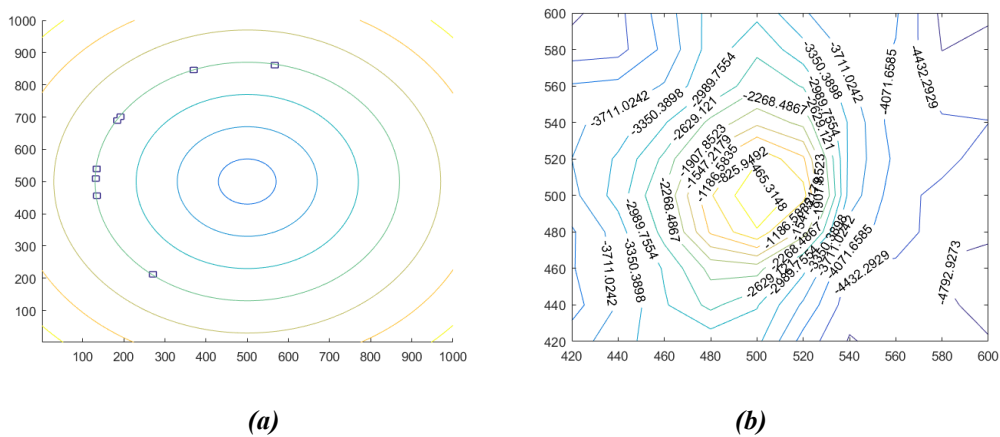

*(a)* *(b)*

*Figure 2 (a) Simulated fire with no wind and homogenous fuel. The concentric circles represent the fire perimeter at various times, increasing from the center outward. The small squares represent simulated satellite fire detections at a time concurrent with the time of the fire perimeter which they overlay. (b) Contour map of data log-likelihood for simulated fire with ignition time $t = 20$. The actual ignition point of the simulated fire was at the point (500,500).*

For this first test, a collection of 500 simulations was made over a 10-by-10 spatial grid at five separate times. The data likelihood of each simulation was then computed and the best estimate of the time and place of ignition was determined using the likelihood function in equation (5). For the set of simulated detections used, the maximum likelihood of all simulations occurred at the correct "true ignition point" but at an incorrect ignition time. The "true ignition time" was $t = 30$, but the estimation procedure gave an earlier time $t = 20$. This is not a surprising result, as an earlier ignition time results in fire perimeter containing more of the area encompassed by the simulated fire detection pixels. As seen by the shape of the likelihood curve in Figure 1 (d), areas within and close to the fire perimeter have a high probability of detection but areas outside of the perimeter at a similar distance may have a very low probability of detection. The earlier ignition time makes for a larger fire, which covers more of the detection squares, which leads to a greater data likelihood.

## 3.2. Test Using the WRF-SFIRE Hill Ideal Experiment

A second test of the data likelihood function to estimate time and place of ignition was made using the WRF-SFIRE coupled atmosphere-wildfire model. This system will be used to work with real fire data but for this second test, simulated fire data created by the "Hill Experiment" within WRF-SFIRE was used. The "Hill Experiment" simulates a fire in a square region with sides of length 2 km and contains a small, dome-shaped hill 100 meters tall situated in its center. A simple atmosphere state, with winds blowing from the northeast was used to create the initial conditions of the weather. Like the first simple test, a simulated fire was created, this time using the WRF-SFIRE model, and then artificial satellite fire detection pixels were created by hand by placing the centers of fire detection pixels on top of a particular fire perimeter. For this simulated fire, the ignition point was chosen to be $(x,y) = (1400\text{m}, 1400\text{m})$ and the ignition time was chosen to be $t = 60$s. Fire detection pixels were placed over the fire perimeter line corresponding to the time $t = 400$s.

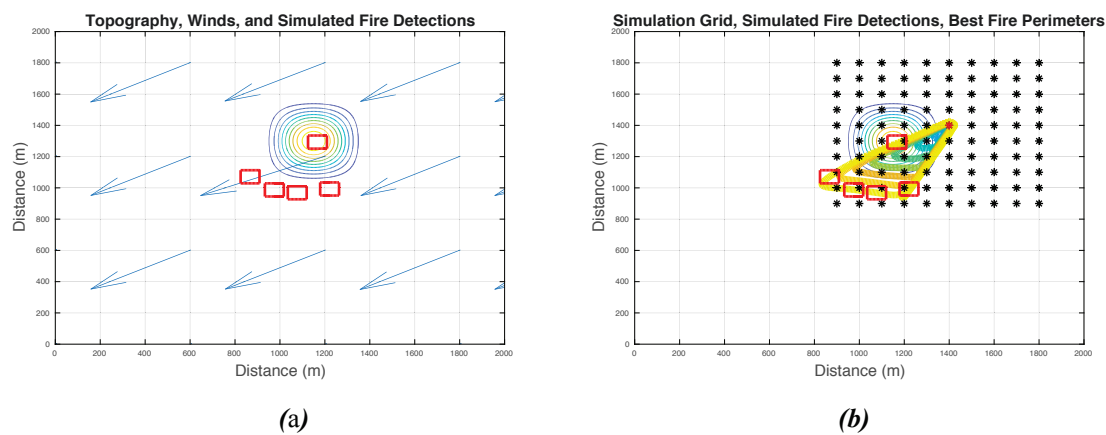

*(a)*                                                                                    *(b)*

***Figure 3** (a) Topography, winds, and simulated fire detection pixels of the "Hill Experiment." The contour lines represent a small hill in the center of a simulated fire domain with winds initially blowing from the upper right corner. The rectangles within the figure are simulated fire detection pixels. (b) Fire simulation grid and fire perimeters of the simulation with largest data likelihood. Fire simulations were started at each of the points on the grid in the figure and the data likelihood of each simulation was then calculated. The colored contour lines represent the fire perimeters of the simulation with the maximum data likelihood.*

Again, a large number of fire simulations were then run at various locations and times and the data likelihood of each was computed to give an estimate of the time and place of the true fire ignition. We ran 300 simulations of the fire on with trial ignition points selected on a 10-by-10 spatial grid, at three distinct ignition times. The true time and place of ignition was in the exact center of this three-dimensional grid of the trial ignition points. The likelihood of the data associated with each simulation was computed and the maximum value found corresponded to the true ignition time and place.

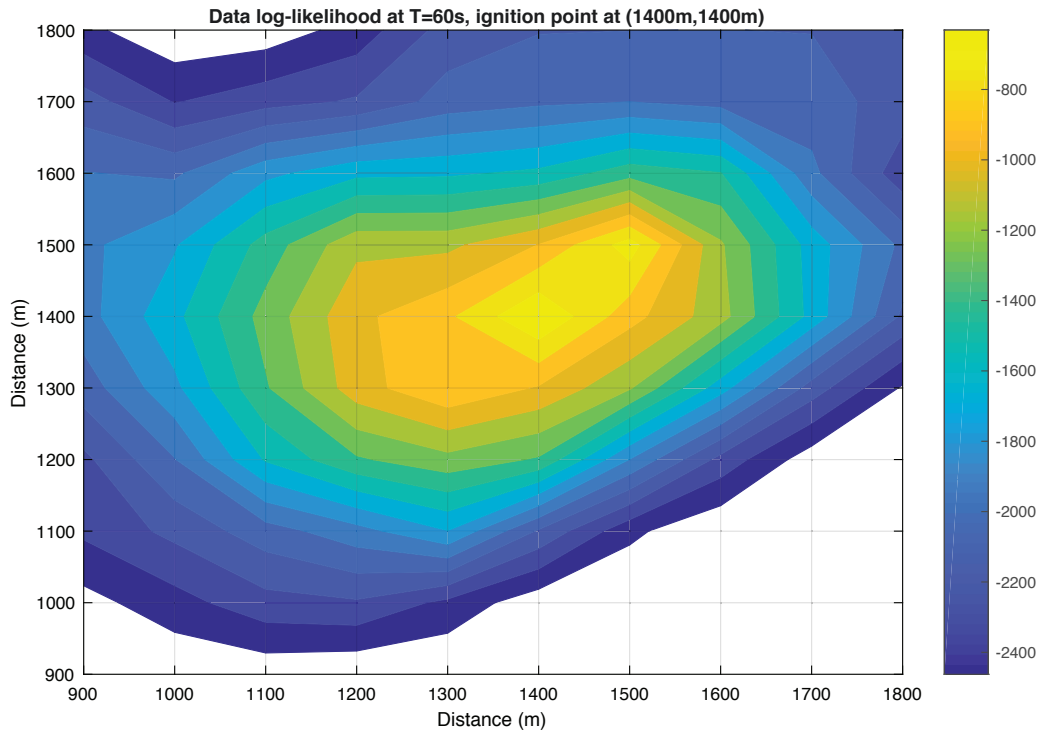

**Figure 4. Contour map of data log-likelihood for ignition points of the WRF_SFIRE Hill Experiment. In this case, the maximum likelihood gives the correct time and location of the fire ignition.**

### 3.3. The 2013 Patch Fire Experiment

A third experiment was completed using satellite data from both MODIS and VIIRS to estimate the time and place of ignition of a real fire. The fire modeled is known as the "Patch Springs Fire" and occurred southwest of Salt Lake City, Utah in August, 2013. As with the other cases previously detailed, the time and place of the fire's ignition was estimated by running an ensemble of WRF-SFIRE simulations at various times and places with the intent to estimate the ignition point by evaluating the data likelihood for each simulation. In this case, 1000 fire simulations were run on a 10-by-10 spatial grid (Figure 5) at ten different ignition times. The spatial grid resolution was approximately 500 meters and the ignition times were spaced two hours apart, with the first ignition time occurring at 21:00 UTC on August 10. The data likelihood of each simulation was evaluated using satellite data from the first two days of the fire simulation period.

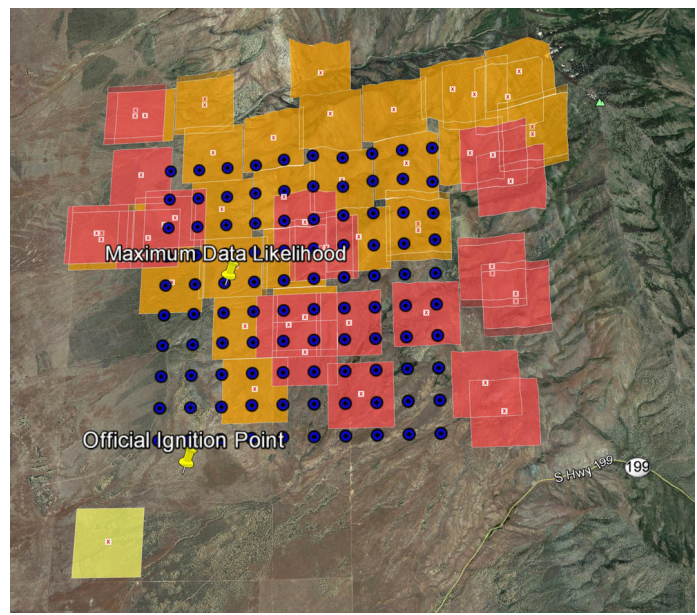

*Figure 5. Satellite fire detections and WRF-SFIRE simulation locations for the Patch Springs Fire of August 2013. The large colored squares represent 1 km active fire locations from the MODIS satellites during the first two days of the fire. The evenly spaced blue points are locations at which fire simulations were run in order to estimate the time and place of the fire ignition. The official estimate of the ignition point lies just outside of the grid of simulations and the estimate of this location obtained by maximum data likelihood lies nearly 3 km to the north.*

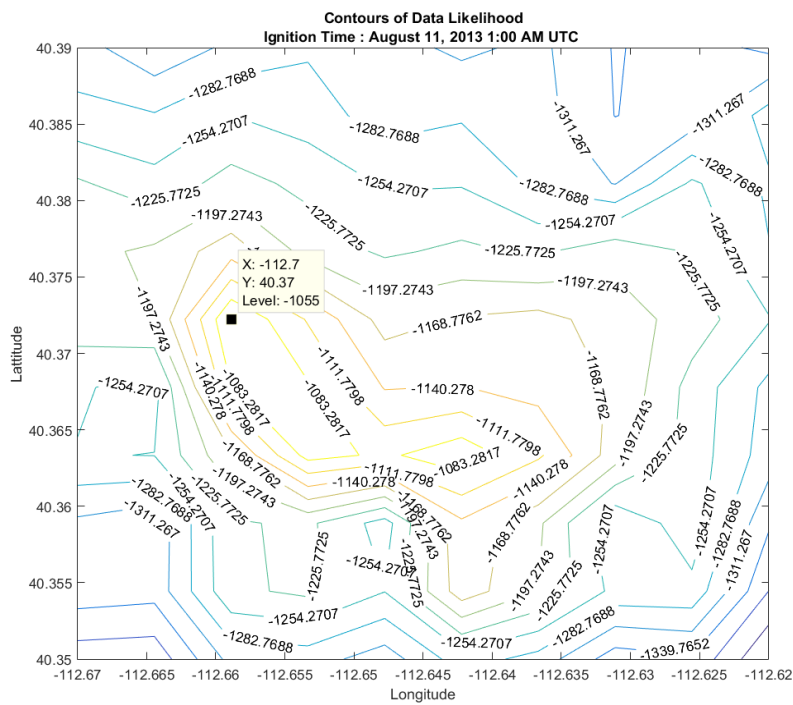

*Figure 6. Contour map of data likelihood for simulations of the Patch Springs Fire on August 11, 2013. The contour lines are drawn from the data likelihood of all fires with simulated ignition times of 1:00 UTC. The best estimate of the fire's ignition point lies just left of center at approximately 40.37°N, -112.7W.*

With all simulations run and the data likelihood of each evaluated, the estimated time and place of ignition was determined to be 40.372°N,-112.659°W at 1:00 UTC on August 11, 2013 (Figure 6). Spatially, this estimate differs by nearly 3 km from the official ignition point determined by investigators. The estimated time of ignition is within an hour of the official ignition time of 2:00 UTC. This fire initially progressed to the northeast but eventually spread southward. It is probable that if that if longer simulations were run and more satellite fire detections showing this southward progression of the fire were used then our estimate would be closer to the official fire ignition location.

## Conclusion

We have derived a physics-motivated likelihood function for active fires satellite detection data, and demonstrated its utility on identifying the ignition point of a wildland fire in time and space. The new data likelihood can be also useful in data assimilation for fire spread models, e.g., as a part of the objective function in an optimization approach (Farguel Caus et al. 2018).

## Acknowledgements

This research was partially supported by grants NSF ICER-1664175 and NASA NNX13AH59G. High-performance computing support at CHPC at the University of Utah and Cheyenne (doi:10.5065/D6RX99HX) at NCAR CISL, sponsored by the NSF, are gratefully acknowledged. We would like to thank Brad Quayle and David Hayes from U.S. Forest Service for supplying the satellite data used in the example in Section 3.3.

## Appendix A

The probability of MODIS fire detection is estimated in validation studies (Hawbaker et al. 2008; Schroeder et al. 2008) by logistic regression as a function of the fire size, or the fraction of the pixel actively burning, as

$$P(d=1) = \frac{1}{1+\exp(-aF+b)}, \tag{1}$$

where $d=1$ means that a fire was detected and $d=0$ that it was not. The constants $a$ and $b$ depend on the properties of the fuel and its loading. They are commonly specified in terms of the false detection rate $1/(1 + \exp(b))$ for $F = 0$, and the quantity $F_{0.5}$, defined by $1/(1 + \exp(-aF_{0.5}+b))=0.5$, which determines $a$ as $a = b/F_{0.5}$. We consider the quantity $F$ in the logistic regression (1) as a proxy for the heat flux $h(x)$ over the pixel at $x$, and thus use (1) with $h(x)$ in place of $F$. This view is supported by the facts that the quantity the hardware sensor actually measures is the radiative power (in its frequency bands); a more accurate regression was obtained in Schroeder et al. (2008) by adding as another regressor the maximum contiguous area burning, which is essentially a proxy for the fire intensity and thus radiative power, which the fine-resolution sensor used for the validation could not measure; and 50% probability of VIIRS detection occurs at nearly constant product of fire size and black body radiation intensity over a range of fire sizes and fire temperatures, as one can compute from Figure 4 in Schroeder et al. (2014).

The heat flux is modelled from the burn model in WRF-SFIRE (Mandel et al., 2011) as identically zero before the fire arrival time $T$ and by a decaying exponential function afterwards,

$$h\left(T\right) = \begin{cases} e^{-(T_{now}-T)/c} & if \quad T_{now} \geq T \\ 0 & if \quad T_{now} < T \end{cases} \qquad (2)$$

where $T_{now}$ is the time of the satellite imaging (Figure 1(a)). However, because of the geolocation error, fire detection in a pixel with nominal coordinates $x = (x_1, x_2)$ is a two-step random process. Assume that the geolocation error is a Gaussian random variable in two dimensions with zero mean and isotropic standard deviation σ. Then, the location $y = (y_1, y_2)$ that the sensor is actually looking at is a random variable with the probability density

$$P(y \mid x) = \frac{1}{2\pi\sigma^2}\exp\left(-\frac{\|x-y\|^2}{2\sigma^2}\right),　\qquad (3)$$

and the probability of detection is given by the logistic formula (1) with $F = h(T(y))$, the fire heat flux at $y$, with $y$ sampled following (3). Assuming further that the geolocation error and the fire detection outcome are independent random variables, the probability of the detection is then given by the probability mixture

$$P(d = 1 \text{ at } x \mid T) = \iint \frac{1}{2\pi\sigma^2}\exp\left(-\frac{\|x-y\|^2}{2\sigma^2}\right)\frac{1}{1+\exp\left(-ah(T(y))+b\right)}dy_1 dy_2 \ . \qquad (4)$$

The probability of no fire detection (i.e., ground without fire) is the complementary probability,

$$P(d = 0 \text{ at } x \mid T) = 1 - P(d = 1 \text{ at } x \mid T)$$
$$= \iint \frac{1}{2\pi\sigma^2}\exp\left(-\frac{\|x-y\|^2}{2\sigma^2}\right)\left(1-\frac{1}{1+\exp\left(-ah(T(y))+b\right)}\right)dy_1 dy_2 \qquad (5)$$

since

$$\iint \frac{1}{2\pi\sigma^2}\exp\left(-\frac{\|x-y\|^2}{2\sigma^2}\right)dy_1 dy_2 = 1 \ .$$

To recover a form of data likelihood similar to that in Mandel et al. (2014b), consider the case of a straight fire line propagating in unit normal direction $n$ with rate of spread $R$. Then,

$$T(y) = T(x) + \frac{\langle y-x, n \rangle}{R} \ . \tag{6}$$

Substituting $T(y)$ into (4), we see that, in particular, the probability of fire detection at $x$ depends on $T(x)$ only. Now take $n = (0,1)$ and $x = (0,0)$, then (6) becomes $T(y) = T(0,0) + y_1/R$, and we have that

$$P(d=1 \text{ at } x \mid T(x)) = \iint \frac{1}{2\pi\sigma^2} \exp\left(-\frac{(x_1 - y_1)^2 + y_2^2}{2\sigma^2}\right) \frac{1}{1 + \exp(-ah(T(x) + \frac{y_1}{R}) + b)} \, dy_1 \, dy_2. \tag{7}$$

The graph of an example of this likelihood function is shown in Figure 1(d), which is indeed similar to the likelihood from Mandel et al. (2014b, 2016b).

For data assimilation and identification of the ignition point, however, we need likelihood computed for the entire satellite image, not just one pixel. First, from (5), we have for both $d = 1$ and $d = 0$ that

$$P(d \text{ at } x \mid T) = \iint \frac{1}{2\pi\sigma^2} \exp\left(-\frac{\|x-y\|^2}{2\sigma^2}\right) \left(1 - d + \frac{2d-1}{1 + \exp\left(-ah(T(y)) + b\right)}\right) dy_1 \, dy_2. \tag{8}$$

So suppose we are given a mesh of pixels at nominal locations $x^i = (x_1^i, \ x_2^i)$ and corresponding detection data $d_i = 0$ or 1, with confidence levels $c_i$ between 0 and 1. Confidence level 0 means missing data. Assume for simplicity that the detections are independent random variables, thus the detection probabilities at individual pixels multiply (and their logarithms add), and discretize the integral in (4) by summation over the same mesh. Then, we can approximate the log likelihood of the detection array $d = (d_i)$ given the array of fire arrival times $T = (T(x^j))$ by

$$\log P(d \mid T) = \sum_i c_i \log \sum_j w_i \exp\left(-\frac{\|x^i - x^j\|^2}{2\sigma^2}\right) \left(1 - d_i + \frac{2d_i - 1}{1 + \exp\left(-ah(T(x^j)) + b\right)}\right), \tag{9}$$

where $w_i$ are normalization constants such that

$$w_i \sum_j \exp\left(-\frac{\|x^i - x^j\|^2}{2\sigma^2}\right) = 1 \ .$$

The data likelihood (9) is used in the examples in Section 3.